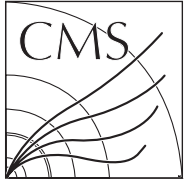

**The Compact Muon Solenoid Experiment**

# Conference Report

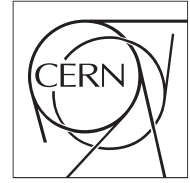

Mailing address: CMS CERN, CH-1211 GENEVA 23, Switzerland

05 January 2009

# The CMS High Level Trigger: Commissioning and First Operation with LHC Beams

M. Felcini, M. Zanetti

## Abstract

The CMS experiment will collect data from the proton-proton collisions delivered by the Large Hadron Collider (LHC) at a centre-of-mass energy up to 14 TeV. The CMS trigger system is designed to cope with unprecedented luminosities and LHC bunch-crossing rates up to 40 MHz. The unique CMS trigger architecture only employs two trigger levels. The Level-1 trigger is implemented using custom electronics. The High Level Trigger is implemented on a large cluster of commercial processors, the Filter Farm. Trigger menus have been developed for detector calibration and for fulfilment of the CMS physics program, at start-up of LHC operations, as well as for operations with higher luminosities. A complete multipurpose trigger menu developed for an early instantaneous luminosity of 1032cm-2s-1 has been tested in the HLT system under realistic online running conditions. The required computing power needed to process with no dead time a maximum HLT input rate of 50 kHz, as expected at startup, has been measured, using the most recent commercially available processors. The Filter Farm has been equipped with 720 such processors, providing a computing power at least a factor two larger than expected to be needed at startup. Results for the commissioning of the full-scale trigger and data acquisition system with cosmic muon runs are reported. The trigger performance during operations with LHC circulating proton beams, delivered in September 2008, is outlined and first results are shown.

# The CMS High Level Trigger: Commissioning and First Operation with LHC Beams

M. Felcini, M. Zanetti
on behalf of the CMS Collaboration

*Abstract*– The CMS experiment will collect data from the proton-proton collisions delivered by the Large Hadron Collider (LHC) at a centre-of-mass energy up to 14 TeV. The CMS trigger system is designed to cope with unprecedented luminosities and LHC bunch-crossing rates up to 40 MHz. The unique CMS trigger architecture only employs two trigger levels. The Level-1 trigger is implemented using custom electronics. The High Level Trigger is implemented on a large cluster of commercial processors, the Filter Farm. Trigger menus have been developed for detector calibration and for fulfilment of the CMS physics program, at start-up of LHC operations, as well as for operations with higher luminosities. A complete multipurpose trigger menu developed for an early instantaneous luminosity of $10^{32}$cm$^{-2}$s$^{-1}$ has been tested in the HLT system under realistic online running conditions. The required computing power needed to process with no dead time a maximum HLT input rate of 50 kHz, as expected at startup, has been measured, using the most recent commercially available processors. The Filter Farm has been equipped with 720 such processors, providing a computing power at least a factor two larger than expected to be needed at startup. Results for the commissioning of the full-scale trigger and data acquisition system with cosmic muon runs are reported. The trigger performance during operations with LHC circulating proton beams, delivered in September 2008, is outlined and first results are shown.

## I. INTRODUCTION

The CMS detector [1] is now built and in its final commissioning phase [2], preparing to collect data from the proton-proton collisions to be delivered by the Large Hadron Collider (LHC), at a centre-of-mass energy of up to 14 TeV. The CMS experiment employs a general-purpose detector with nearly complete solid-angle coverage, which can efficiently and precisely measure electrons, photons, muons, jets (including tau- and b-jets) and missing energy over a wide range of particle energies and event topologies. These characteristics ensure the capability of CMS to cover a broad programme of precise measurements of Standard Model physics and discoveries of new physics phenomena. The trigger and data acquisition system must ensure high data recording efficiency for a vast variety of physics objects and event topologies, while applying online very selective requirements.

The CMS trigger and data acquisition system [3, 4] is designed to cope with unprecedented luminosities and interaction rates. At the LHC design instantaneous luminosity of $10^{34}$cm$^{-2}$s$^{-1}$ and bunch-crossing rates of 40 MHz, an average of about 20 interactions will take place at each bunch crossing. The trigger system must reduce the bunch crossing rate to a final output rate of O(100) Hz, consistent with an archival storage capability of O(100) MB/s.

The trigger menus (the sets of the reconstruction and filtering algorithms and their configuration) must be chosen and optimized to address the detector needs and physics objectives of the experiment, depending on luminosity, machine and detector conditions. According to the LHC start-up plan, the LHC instantaneous luminosity (hereafter referred to as luminosity L), in the initial phase, is expected to increase gradually before reaching the design luminosity. Runs at low luminosities will be useful to fully commission and calibrate the detector as well as to measure Standard Model processes, before reaching the high luminosity phase, when discoveries of new physics phenomena will be the main goal of the experiment.

In this paper, after a concise description of the CMS trigger and data acquisition (DAQ) system, we outline the tests performed for the High Level Trigger (HLT) deployment and validation in the online environment, commissioning with cosmic muon data and first operations with LHC circulating beams.

## II. TRIGGER AND DAQ SYSTEM

The CMS trigger architecture employs only two trigger levels. The Level-1 Trigger (L1T) [3] is implemented using custom electronics. The HLT [4] is implemented on a large cluster of commercial processors, the HLT Filter Farm. The architecture of the CMS Trigger and DAQ system is shown schematically in Fig. 1.

The L1T system must process information from the CMS detector at the full bunch crossing rate (up to 40 MHz at the highest LHC luminosities). The time between two successive bunch crossings, along with the wide geographical distribution of the electronic signals from the CMS sub-detectors, require the use of fast electronics. The time for processing the detector information in the L1T system is limited by the front-end (FE) electronics capability to store the detector data during the L1T decision process. The FE electronics modules can store the data from up to 128 contiguous bunch crossings, i.e. ~3 μs. Within this time interval, the detector information must be transferred to the L1T processing elements, the decision must be formed and the decision signal must be transferred back to the FE electronics. The resulting time available for processing the data in the L1T system is no more than ~1μs. Thus the L1T can process a limited amount of detector data, from calorimeters and muon chambers, with coarser granularity and lower resolution than the full information recorded in the FE electronics. The processing elements of the L1T system are custom-designed. Details of the architecture, the design and the selection algorithms in the L1T can be found in [3]. The L1T system is designed to achieve a bunch crossing rate reduction factor of up to 400, for a maximum mean event accept rate of 100 kHz. The estimated average size of an event record is O(1MB). After the acceptance of an event by the L1T, about 700 FE modules store the event data, each carrying 1-2 kB of data per L1T accepted event.

The next online selection step, the HLT [4], must operate a rate reduction factor of ~1000, dictated by the ability to store and reconstruct data offline at a maximum accept rate of O(100) Hz, or O(100) MB/s. Such a rejection factor requires that the HLT selection be based on full granularity and resolution information from the whole detector, including trackers, with reconstruction algorithms almost as sophisticated as those used in the offline event reconstruction. This implies the usage of fully programmable commercial processors for the execution of the HLT. The expectation that the HLT algorithms will demand a mean processing time of O(10) ms, along with the maximum HLT input rate of 100 kHz, implies that O(1000) processors in the filter farm must be employed for this processing stage. This, in turn, requires the Data Acquisition (DAQ) system [4] to provide the means to feed data from ~700 FE modules to about 1000 processors, at a sustained bandwidth of up to 100 kHz×1MB=100 GB/s. The interconnection of such a large number of elements, at such a bandwidth, implies the usage of a switching network (Builder Network). Two systems complement this flow of data from the FE memories to the Filter Farm: the Event Manager, responsible for the actual data flow through the DAQ, and the Control and Monitor System, responsible for the configuration, control and monitor of all the elements.

The HLT/DAQ data flow is illustrated in Fig. 2. For each L1 accept, event fragments are readout and stored in readout buffers called Readout Units (RU). Event fragments are then assembled into complete events by the Event Builder Network. Complete events are handed to Builder Units (BU) feeding the HLT Filter Units (FU) which process the event content to form the HLT decision. Several BU+FU (BUFU) nodes (Fig. 2) process events in parallel. Events accepted by the HLT are forwarded to the Storage Manager, which streams event raw data on disk and subsequently transfers raw data files to the CMS Tier-0 computing center at CERN for permanent storage and offline processing.

III TRIGGER MENUS FOR LHC START-UP LUMINOSITIES

During the first year of LHC operation, the CMS DAQ system is set up to be able to sustain an event readout rate of up to 50 kHz from the L1T system. Events processed by the Filter Farm, running HLT reconstruction and selection algorithms, will be accepted at a rate of a few hundreds Hz for output to permanent storage.

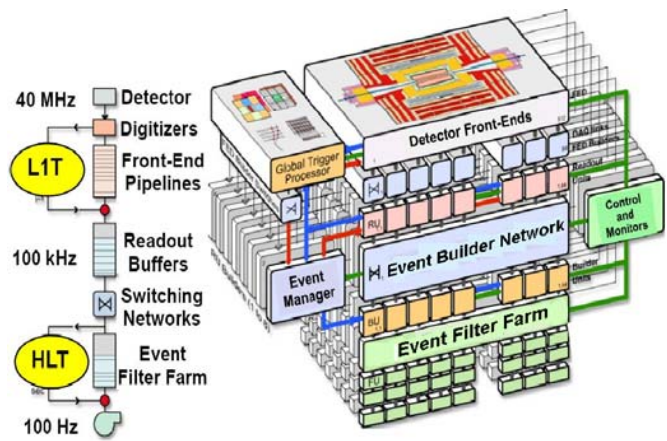

**Figure 1**. Schematic view of the CMS trigger and DAQ system showing, left, the successive stages and, right, the modularity (slices) of the system [4].

In the following, we recall the motivation and the results of a detailed study **[5]** aimed to develop HLT menus with the expected physics and computing performance for LHC startup luminosities up to O $(10^{32})$ cm$^{-2}$s$^{-1}$.

The actual trigger performance will be measured and optimized with real data from collisions, when the actual experimental (collider and detector) conditions will be known. We can use our present best knowledge of the detector response and possible collider condition scenarios to study and optimize trigger criteria, in view of adjusting them when real collisions and detector data will be available. The flexibility of the trigger system allows to introduce modifications in an efficient manner for optimal performance, adapted to the actual running conditions while taking data. The robustness of the algorithms which determine the trigger primitives also ensure that the system will not be too sensitive to small changes with respect to the expected conditions.

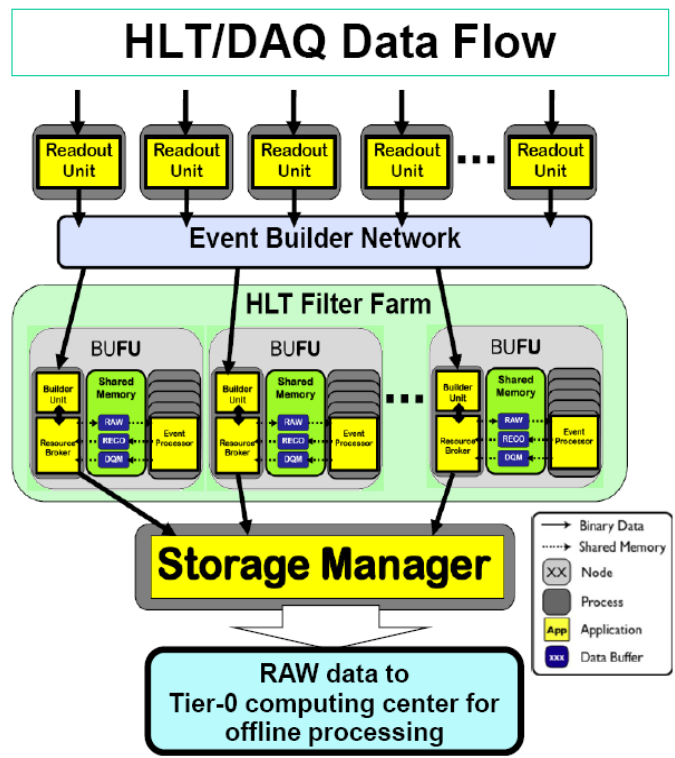

Figure 2. HLT/DAQ data flow of L1 accepted events, through the HLT Filter Farm, up to offline permanent storage and processing at Tier-0.

In preparation for data taking, the HLT menus are developed using fully simulated and reconstructed Monte Carlo events for the known Standard Model processes, dominated in rate by QCD events at low transverse momentum, also known as minimum bias events. The goal is to achieve a reduction factor of ~1000 on the input HLT rate, while keeping as high as possible efficiency for events of interest, with an average processing time per event of the single HLT instance not exceeding ~50 ms [4,5]. To meet these requests, the menus are structured as a set of trigger paths, each of which is dedicated to select events with specific topologies and kinematics. The selection can be inclusive, requiring the presence of single or double physics objects e.g. muons, photons, jets, etc., or exclusive, for events featuring particular final states with multiple physics objects (cross-triggers).

The CPU time required for the execution of the of the HLT algorithms in the filter farm is minimized by rejecting events as quickly as possible, using the minimum amount of detector information. In each path the filtering modules are embedded among the reconstruction sequences so that if the filter requirements are not matched, the rest of the path is not executed. Identical instances of the same algorithm present in several paths are executed only once. In addition only partial event reconstruction is performed on those parts which can be used for immediate selection. After each possible reconstruction step, a set of selection criteria, applied to the reconstructed objects, results in the rejection of a significant fraction of events, thus minimizing the CPU usage at the next step.

In the summer 2007, a complete HLT menu with these features has been produced [5] as the candidate menu for physics runs at $L=10^{32}$ cm$^{-2}$s$^{-1}$.

III HLT ONLINE VALIDATION

In order to test and validate candidate HLT menus in realistic online conditions [6, 7] and estimate the required computing power, a playback system was put in place and used as a test bed.

The playback setup, shown schematically in Fig. 3, uses the same architecture of the BUFU nodes as in the full DAQ system, but in replacement of the events provided by the Event Builder Network (see Fig. 2), an application reads the events from a data file and passes them to an ad-hoc version of the BU (auto-BU). From this level on, the data exchange protocol is the same as described in Section II. The auto-BU is configurable in such a way that once event data are read from the file, they can be re-played continuously.

Several BUFU nodes can be run in parallel, having one or more storage manager receiving events from each of them. This allows to test the scalability of the system as well as to compare simultaneously the performances of different commercially available CPUs. For the validation of the HLT

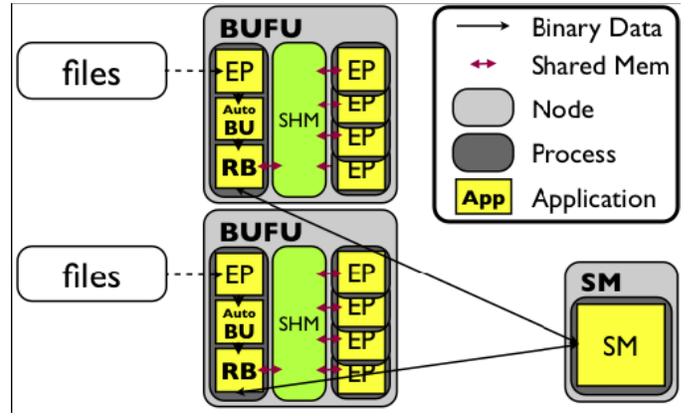

Figure 3. Layout of the playback setup of the Filter Farm used for the online validation of the HLT code and configurations. Events are readout from Monte Carlo event files, by means of a dedicated application replacing the Event Builder Network. Then they are passed to the Filter Units using the standard data exchange protocol

menu [5] at $L=10^{32}$ cm$^{-2}$s$^{-1}$, 20 BUFU nodes have been employed. The frequency of the CPU's installed in those nodes ranged between 2 and 3 GHz.

As input for the validation tests, a Monte Carlo sample of 20 millions minimum bias events was produced and passed through the full detector simulation. The L1 menu decision was emulated in order to select only those events which would pass the L1 trigger, reducing the initial sample to ~$10^7$ events. Realistic detector conditions were simulated, corresponding to uncertainties on the calibration constants as expected after collecting an integrated luminosity of 100 pb$^{-1}$. The distribution of the event processing time obtained by running the HLT menu on this event sample is shown on Fig. 4. The mean processing time per event is ~43 ms and matches well the initial design requirements. This result is obtained on a 2.66 GHz CPU.

The scaling properties of the HLT processing time as a function of the CPU frequency have been studied. The average number of events processed per second as a function of the processor frequency in GHz is shown in Fig. 5, obtained by comparing four Intel® processors. This result demonstrates the linear increase of the CPU performance with the processor frequency.

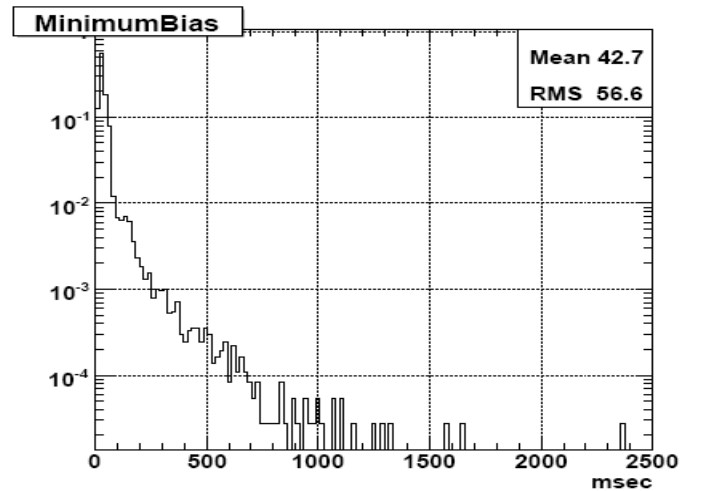

Figure 4. CPU time distribution as obtained when running the HLT menu on minimum bias simulated events, accepted by the L1 emulator. This result is obtained on a dual-core 2.66 GHz CPU PC.

Recently, commercially available processors combine two or more independent cores into a single die, or more dies packaged together. The performance of multi-core processors running several instances of the HLT executable has been tested. Fig. 6 shows the number of processed events per second as a function of the number of HLT instances running on a dual quad-core machine. The number of processed events increases almost linearly up to eight instances, with each of the single processes using constantly ~100% of the available CPUs[1]. When nine, or more, processes run in parallel the overall CPU power is equally shared among each process, yielding the same timing result achieved with eight processes.

The memory usage could be an issue in the case of multiple HLT instances running concurrently on the same node. The HLT menu under test was measured to demand not more that 600 MB of resident memory. A memory slot as large as 2 GB was assigned for each core.

The power consumption results not to be particularly affected by multiple processes running in parallel on the same node. The increase of the power consumption when all the CPU's are fully used with respect to the idle state is ~20%.

Given the results of the tests on the candidate HLT menu for $L=10^{32}$ cm$^{-2}$s$^{-1}$, it has been decided to base the Filter Farm on PC's mounting 2.66 GHz dual quad-core Intel® processors (*Clovertown*) 16 GB RAM. In October 2008 these PC's were installed in the CMS counting room, commissioned and employed to run the HLT menus during CMS global data taking runs with cosmic muons (see Section IV).

Assuming a conservative factor of two in the average HLT processing time, corresponding to an average number of ~100 events processed per second and per core, the CPU power currently available in the Filter Farm will be suitable for handling an HLT input rate as high as ~60 kHz, beyond what expected for the first year of LHC physics runs. An additional safety factor is provided by the fact that more recently developed HLT menus reduce the average processing time to less than 30 ms/event. Moreover, the computing power of the CMS Filter Farm is expected to be doubled in size, with more recent CPU's, before startup of LHC operations at high luminosity.

## IV. HLT COMMISSIONING WITH COSMIC RAYS

Since the final assembly of the detector in the underground cavern in summer 2007, a constantly increasing fraction of the CMS sub-detectors are being commissioned and integrated with the trigger and DAQ systems. Global data taking campaigns recording cosmic muon events are taking place regularly with the goal to commission the experiment for the first beam data. In this context, the HLT system plays an important role, providing selected data sample for commissioning of the different sub-detectors and linking the online environment with the offline computing and analysis workflows. The commissioning of the HLT system itself implies the deployment of online menus of increasing complexity.

---

[1] The other two processes running on the BUFU node, i.e. the (auto)BU and the RB require very little CPU power, less than 5% in total.

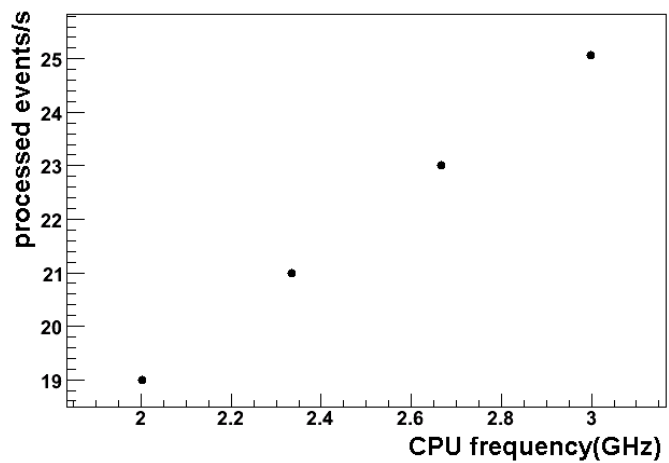

Figure 5. Number of processed events per second per HLT instance as a function of the CPU frequency. All four measurement are done on Intel processors

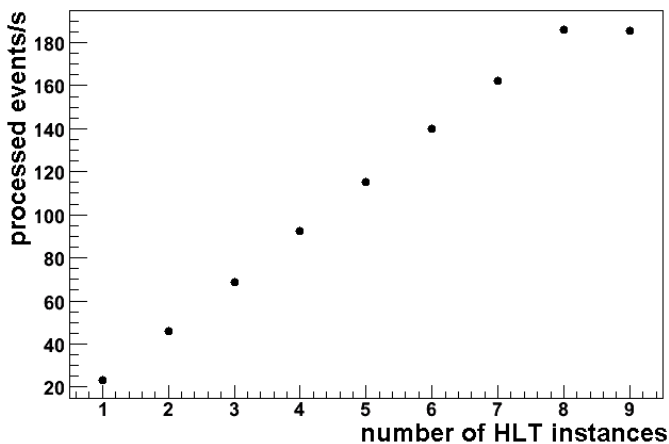

Figure 6. Number of processed events per second as a function of the number of HLT instances running on a 2.66 GHz dual quad-core machine.

### A. HLT configurations for cosmic ray data taking

During commissioning with cosmic muons, ad-hoc code and menu configurations have been employed in the HLT. For example, at least one path in the HLT menu is set to accept all physics events, having passed the L1 trigger, while other HLT paths apply very simple filters (based on the reconstructed muon direction or on the number of reconstructed tracks in the tracking system) and tag the events for offline processing and skimming.

The rate of cosmic events detected by CMS does not exceed ~500 Hz. In order to stress the HLT system with a high input rate, Poisson distributed random triggers have been generated by the L1 Global Trigger and delivered together with the cosmic triggers at frequencies as high as 60 kHz. Given the reduced capability of the Filter Farm in terms of computing power during the early detector commissioning phase (~300 CPUs employed), the random triggered events have not been fully reconstructed, but only tagged and streamed out separately from the cosmic muon events.

The complete menu developed for LHC collisions, has also been tested online during cosmic muon data taking. This menu consists of ~100 HLT paths, with more than 1000 different algorithm instances. It has been run steadily online for several days. Given the extremely simply topology of the events, the overall processing time per event has been measured to be about 20 ms.

The first step of the HLT data processing is the unpacking of raw data. The algorithms performing this operation have to be robust against data corruption and missing information and at the same time monitor and log possible readout errors. The reconstruction of physics objects at the HLT starts from the corresponding candidates identified by the L1 trigger. Only the parts of the detector pointed to (seeded) by the L1 information need to be considered for further validation of the trigger object. During commissioning with cosmic rays, the HLT data unpacking and path seeding algorithms have been fully tested, requiring that every HLT path be seeded by on one or more L1 bits. Both muon and calorimeter (jets and $e/\gamma$) L1 bits have been employed. In particular, these tests have demonstrated that L1 muon candidates provide seeds for high efficiency HLT muon reconstruction. The event display in **Fig. 7** shows a muon track reconstructed online by the HLT algorithms, seeded by the information of the L1 Global Muon Trigger. Dedicated muon reconstruction algorithms are implemented in the HLT menus for cosmic muon, to deal with particle trajectories not originating from the center of the detector (interaction point) and to reconstruct the muon trajectory as a single track.

### B. HLT output: streams and primary datasets

The reconstruction software framework requires the output files to be homogeneous in terms of the information content of the events. The need for different HLT data streams arises therefore when different event content has to be used for different data samples. As an example, the event content of the data samples to be used for the offline physics analysis (physics stream) consists normally of the complete collection of detector and trigger raw data, as well as of the L1 and HLT selection results. On the other hand, detector calibration can benefit form the highest possible collection rate. Hence, calibration and alignment streams only store a portion of the raw data or dedicated collections of reconstructed objects, allowing higher HLT accept rates than the physics streams.

The concurrent production of several data streams has also been tested during the commissioning runs with cosmic rays. In particular, alignment and calibration ("AlCaRaw") streams, as foreseen for LHC collision data taking, have been added to the main physics stream.

Within a stream, sets of paths performing similar selections can be further grouped to define primary datasets. The definition of the streams and of the primary datasets is part of the HLT configuration and has important implications for the offline computing operations. The data stream files, produced by the HLT and written out by the Storage Manager, are transferred to the CMS Tier-0 computing center at CERN, where each stream is processed accordingly to its predefined workflow. AlCaRaw streams are processed first, in order to provide calibration and alignment constants to be used for the reconstruction of the events in the physics stream. In the offline processing, the physics stream is split into primary datasets, as defined by the HLT configuration. The complete data workflow, from the online environment to the offline reconstruction and analysis, has been tested and used for offline data production, during cosmic ray data taking.

A dedicated data stream ("HLTdebug"), including the collections of reconstructed objects the HLT selections are based upon, is also defined to allow HLT performance monitoring and debugging.

### C. Data Quality Monitor

The online data quality monitoring (DQM) is an important tool for commissioning and efficient data taking, fundamental for the identification and the debugging of issues regarding both the various detector components and the trigger system.

In the online environment, the DQM is performed at two stages: as part of the HLT execution and parasitically from a low rate stream fed by the Storage Manager. Both these modes of operation have been established since the first phases of commissioning with cosmic muon data. In the former case, monitoring applications have access to all L1-accepted events as read out from the detector front ends and while being processed by the different HLT modules. This is the most suitable context to execute data integrity checks and to monitor the performance of the L1 and HLT operations (see Fig. 8 for an example of an HLT DQM result). The DQM results (typically in the form of histograms) are sent by each FU to the Storage Manager, where they are collated and delivered to a web based graphical user interface (GUI). Monitoring operations in the HLT should however be limited to what is strictly needed, in order to interfere minimally with the HLT event reconstruction and selection.

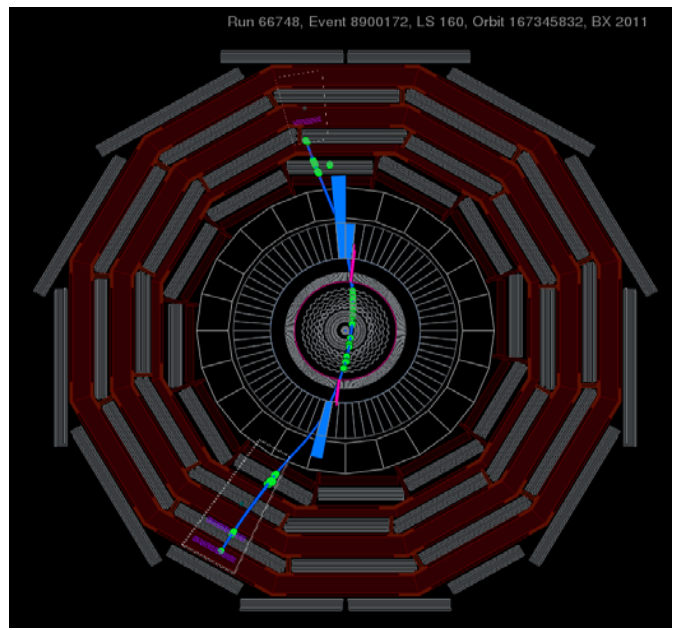

**Figure 7**. Event display of a cosmic muon reconstructed online in the HLT. Information from the muon chambers, calorimeters (m.i.p energy deposits) and tracking system are used to fit the muon track throughout the full detector.

Many of the online DQM applications, dedicated to the monitoring of the detector data, are fed by a proxy server connected to the storage manager, which provides events sampled among all those accepted by the HLT. The DQM applications receive sampled events from the HLTdebug stream, for a predefined subset of the HLT paths. DQM applications can then operate further selections, e.g. displaying monitoring results only for events accepted by a specific HLT path.

## V HLT OPERATION WITH THE FIRST LHC BEAMS

In summer 2008, the accelerator team has started injecting 450 GeV proton beams into the main LHC ring. CMS has detected the first beam induced activity at the beginning of September 2008, when a beam has been steered onto collimators placed upstream of the CMS detector. On September 10, 2008, and in the following nine days, two counter rotating proton beams have circulated (one at the time) along the entire LHC ring, passing several times through CMS. Due to the unfortunate incident occurred in the sector 3-4 of the LHC, the collider operations have been suspended on September 19, 2008, before achieving the first proton-proton collisions. LHC operations will be resumed in 2009, after an extended winter shutdown.

The detection of the first LHC beam induced events [8] has been very useful for the commissioning of the whole experiment, in particular to study the timing of the trigger system and of the various sub-detectors with respect to the beam signal. The trigger and DAQ system has been configured to record single beam events. Two types of L1 triggers have been employed:

- beam halo triggers, used to trigger on the detector activity induced by the beam, such as beam halo muons passing through the muon end-cap (ME) chambers and through the forward hadron calorimeters (HF). The muon halo trigger rates were between ~1 and ~40 Hz (see Fig. 9);
- beam capture triggers, based on the signals from beam detectors, namely Beam Scintillator Counters (BSC) and the Beam Pickup (BPTX), placed along the beam pipe, at various distances from the CMS detector. These devices were providing a trigger signal at every beam passage with a trigger rate as high as ~11 kHz in the case of a (single) continuously circulating beams.

The HLT system was configured to apply no rejection, but simply "tag-and-pass" the events accordingly to the L1 information. In the case of continuously circulating beam, the beam capture trigger was prescaled directly at L1, thus reducing the HLT rate to a level sustainable by the storage system. Each of the paths implemented in the HLT menu was seeded by a specific L1 bit. Thus, the HLT unpacking of the raw data was performed on each L1 accepted event.

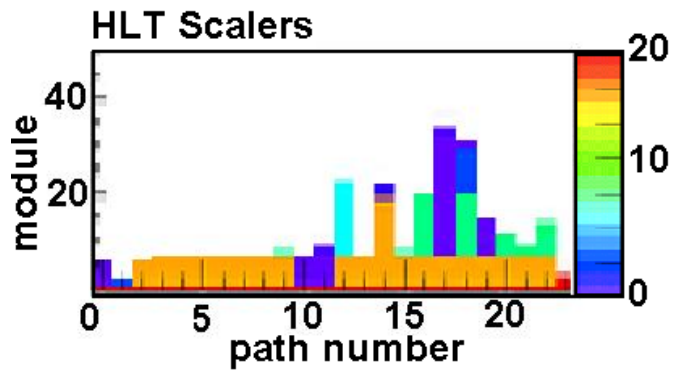

Figure 8 Example of DQM histogram for the monitoring of the HLT performances. The HLT paths are listed on the horizontal axis, whereas the vertical axis represents the depth along the reconstruction sequence. The color code expresses the number of times a given reconstruction module is executed. This result is for a cosmic muon run with dedicated HLT menu.

Beam capture L1 triggers were seeding also two dedicated HLT paths, applying special filters to detect beam-induced signals in the central detectors. These filters were based on very simple quantities, such as the size of the data payload of the pixel vertex detector and the energy deposition measured by the calorimeters. These filters required no reconstruction to be performed, but for the calorimeter local reconstruction.

The data taking was flawless and efficient during the entire period of beam runs. The BPTX signal was synchronized with the trigger system almost immediately (at the third beam shot to CMS). From that moment on, the events have been continuously logged on to disk. Beam events selected by the HLT were immediately made available to the online DQM and the Event Display. The histograms produced in the Filter Farm by the DQM applications have proven very useful to monitor the trigger performances and their evolution with the time. Examples of online event displays are reported in Fig. 10 and 11. The event display in Fig. 10 shows the effect of the beam smashing on the collimators located upstream of the CMS detector: hundreds of thousands of particles passing through the detector deposit a large amount of total energy (up to ~1000 TeV) in the hadron calorimeter. The event display in Fig. 11 shows a beam halo muon, crossing CMS from one end-cap side to the other, and reconstructed in the muon chambers of both end-cap disks.

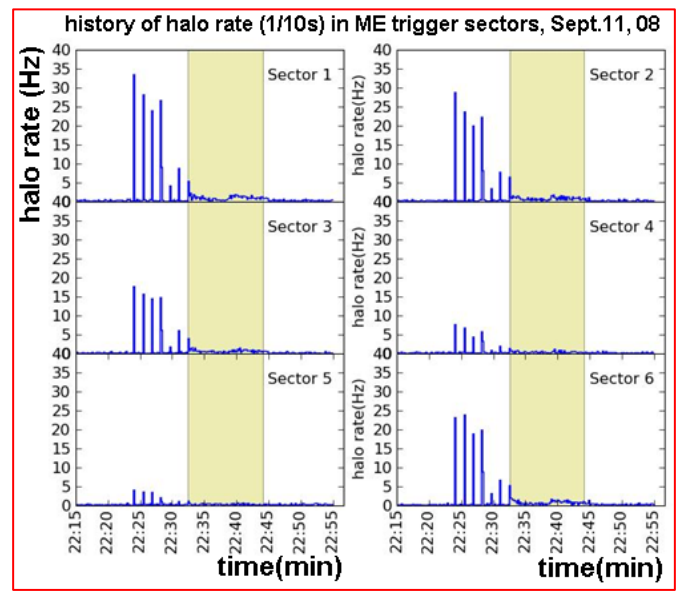

Figure 9. Muon halo rates recorded in the muon end-cap (ME) trigger sectors, at the passage of the beam through CMS. The yellow band shows the beam capture signal recorded downstream of the CMS detector.

## VI SUMMARY AND OUTLOOK

The CMS experiment will collect data from the proton-proton collisions delivered by the Large Hadron Collider (LHC) at a centre-of-mass energy of up to 14 TeV, starting operations in 2009. The CMS trigger system is designed to cope with unprecedented luminosities and LHC bunch-crossing rates up to 40 MHz. The unique CMS trigger architecture only employs two trigger levels. The L1T, implemented using custom electronics, inspects events at the full bunch-crossing rate, while selecting up to 100 kHz for further processing. The HLT reduces the 100 kHz input stream to O(100) Hz of events written to permanent storage. The HLT system consists of a large cluster of commercial processors, the Filter Farm, running reconstruction and selection algorithms on fully assembled event information. L1 and HLT menus have been developed for startup, low luminosity conditions. A total DAQ readout capability of 50 kHz is required at startup. Fast selection and high efficiency is obtained for the physics objects and processes of interest using inclusive selection criteria. The overall CPU power requirement for realistic HLT menus is measured to be within the system capabilities. The results of trigger commissioning with cosmic muon data and first operation with LHC circulating beam demonstrate that the CMS experiment is ready to collect data with high efficiency from the start-up of the LHC operations with colliding proton beams.

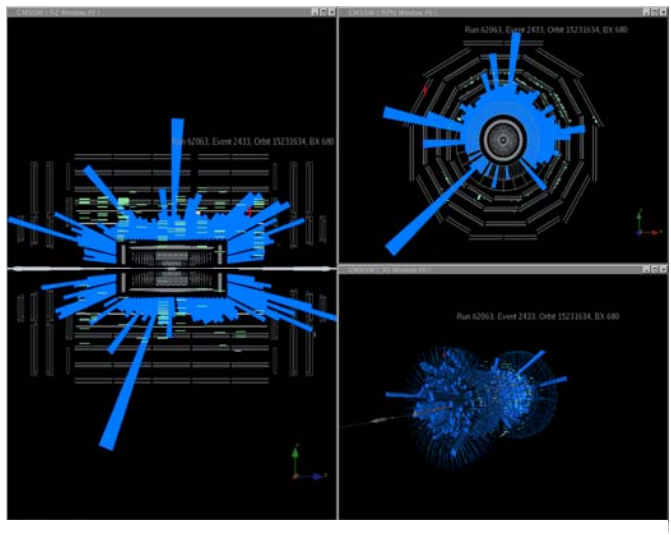

Figure 10. Beam splash event: muons originated from ($2\times10^9$) protons hitting the collimator blocks situated about 150 m upstream of CMS, were detected in the calorimeters (calorimeter deposits shown in blue) and muon chambers (muon chamber hits shown in green).

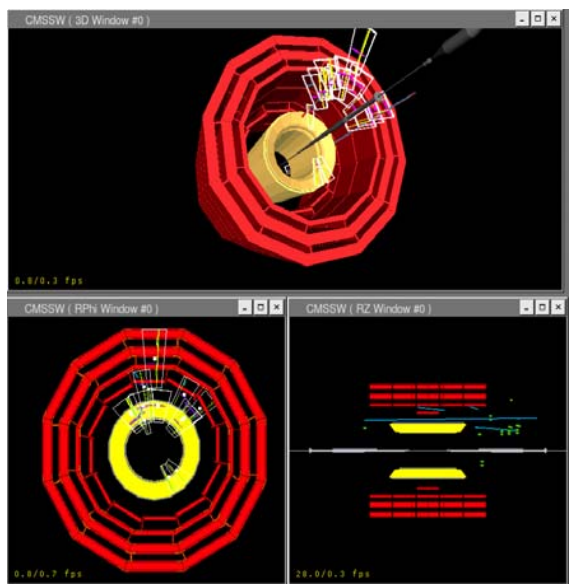

Figure 11. Sections of the CMS detector displaying a beam halo muon.

## ACKNOWLEDGMENT

We are very grateful to our colleagues in the CMS Collaboration and in particular in the Trigger and DAQ groups for giving us the opportunity to present these recent results on their behalf. We wish to thank E. Meschi for careful reading of the manuscript and valuable comments.